\documentclass{article}%
\usepackage{amsfonts}
\usepackage{amsmath}
\usepackage{amssymb}
\usepackage{graphicx}%
\begin{document}

\title{Mean first-passage time of quantum transition processes}
\author{Rong-Tao Qiu$^{1}$, Wu-Sheng Dai$^{1,2}$\thanks{Email: daiwusheng@tju.edu.cn}
and Mi Xie$^{1,2}$\thanks{Email: xiemi@tju.edu.cn}\\{\footnotesize 1) Department of Physics, Tianjin University, Tianjin 300072,
P. R. China }\\{\footnotesize 2) LiuHui Center for Applied Mathematics, Nankai University \&
Tianjin University,} {\footnotesize Tianjin 300072, P. R. China}}
\date{}
\maketitle

\begin{abstract}
In this paper, we consider the problem of mean first-passage time
(MFPT) in quantum mechanics; the MFPT is the average time of the
transition from a given initial state, passing through some
intermediate states, to a given final state for the first time. We
apply the method developed in statistical mechanics for
calculating the MFPT of random walks to calculate the MFPT of a
transition process. As applications, we (1) calculate the MFPT for
multiple-state systems, (2) discuss transition processes occurring
in an environment background, (3) consider a roundabout transition
in a hydrogen atom, and (4) apply the approach to laser theory.

\end{abstract}

\vskip 1cm


\section{Introduction}

In statistical mechanics, for a random-walk process, the mean first-passage
time (MFPT) is the average time for the walker to reach some given final site
for the first time assuming that the walker started at a given initial site
\cite{1,2}.

In quantum mechanics, we can also consider the problem of the MFPT. Like that
in statistical mechanics, the MFPT in quantum mechanics is the average time
for a quantum system to transit from a given initial state at time $t=0$ to a
given final state for the first time.

Concretely, for a quantum transition process $\left\vert i\right\rangle
\rightarrow\left\vert \text{\textrm{intermediate} \textrm{states}%
}\right\rangle \rightarrow\left\vert f\right\rangle $, the MFPT represents the
average time of the transition from the initial state $\left\vert
i\right\rangle $ to the final state $\left\vert f\right\rangle $ for the first
time, where $\left\vert \text{\textrm{intermediate} \textrm{states}%
}\right\rangle $ denotes the intermediate states that the transition process
passing through. Clearly, the MFPT of a transition process contains the
information of all states that the transition passes through. Especially, for
a two-state system, the MFPT of the transition process $\left\vert
i\right\rangle \rightarrow\left\vert f\right\rangle $ is just the lifetime of
the state $\left\vert i\right\rangle $.

The difference between the lifetime and the MFPT is as follows. In quantum
mechanics, what one considers more frequently is the lifetime of a state. When
considering the lifetime of a state, one considers the transition process of a
given initial state to all possible final states, while, when considering the
MFPT, what one considers is a given transition process whose initial, final,
and all intermediate states are assigned.

The MFPT of a quantum transition can be calculated directly by the master
equation. The reason why we use the master equation for the treatment of a
quantum transition process is that a transition occurring in a quantum system
must need external disturbances. In other words, a system which has no
interaction with the environment will not display transitions; or, an isolated
system will keep in the initial state forever. It is just the interaction
between a quantum system and the environment that causes the quantum
transition, e.g., the spontaneous radiation is caused by vacuum fluctuations.
That is to say, a quantum system which can display quantum transitions must be
an open system. Meanwhile, in fact, no real isolated system exists because of
the existence of vacuum fluctuations. Therefore, when we consider a problem of
quantum transition, we will face an open system, and, then, the master
equation comes in handy.

The method we used in the present paper for the calculation of the MFPT is the
master equation method developed in statistical mechanics for the calculation
of the MFPT of a random-walk problem \cite{1}. In this paper, we directly
apply this method to quantum transition problems. In the following, we first
review the method. Then, we calculate the MFPT for multiple-state systems.
Concretely, we calculate the MFPT for two-, three-, four-, and five-state
systems as examples.

As an application, we consider a transition process occurring in an
environment background. In environment backgrounds, the transition will be
disturbed. There are many theretical reaserches on quantum information and
quantum computation \cite{CWKO,DWCO}. Nevertheless, when realizing quantum
information and quantum computation in experiments, one inevitably encounters
the problem of decoherence coursed by environment backgrounds.
\cite{BGA,Per,DLDVT,CFL}.

Moreover, we also consider a roundabout transition in a hydrogen atom.
Concretely, even if a transition is forbidden by selection rules, a roundabout
transition can still occur through an indirect way. For example, if the direct
transition $\left\vert i\right\rangle \rightarrow\left\vert f\right\rangle $
is forbidden, however, a roundabout transition $\left\vert i\right\rangle
\rightarrow\left\vert \text{\textrm{intermediate} \textrm{states}%
}\right\rangle \rightarrow\left\vert f\right\rangle $ can still occur, so long
as the transitions $\left\vert i\right\rangle \rightarrow\left\vert
\text{\textrm{intermediate} \textrm{states}}\right\rangle $ and $\left\vert
\text{\textrm{intermediate} \textrm{states}}\right\rangle \rightarrow
\left\vert f\right\rangle $ are permitted.

Finally, we discuss the application of the MFPT in laser theory.

There are many researches on the calculation and also the application of the
MFPT. The MFPT of the Markov process has been considered in Ref. \cite{5}, and
that of the Non-Markov process has been considered in Refs. \cite{6,7}. The
applications of the MFPT in magnetics have been considered in Ref. \cite{9}.
There are also many studies on the MFPT with potentials or noises
\cite{10,13,15}. The problem of laser with the MFPT has been discussed in Ref.
\cite{17}. The MFPT on T-graph has been considered in Refs. \cite{18,19} and
that in random environments has been considered in Refs. \cite{21,23}. Many
other researches consider the application of the MFPT in other fields, e.g.,
in chemistry \cite{24,27,28} and biology \cite{24}.

In section 2, we apply the master equation method developed in statistical
mechanics to the calculation of the MFPT of quantum transition processes. In
section 3, we calculate the MFPT for multiple-state systems. In section 4, we
consider the problem of transition processes occurring in environment
backgrounds. In section 5, we discuss the problem of roundabout transitions
for a hydrogen atom as an example. In section 6, we consider an application in
laser theory. In section 7, conclusions and outlook are given.

\section{Mean first-passage time of a quantum transition: the master equation
\label{MFPT}}

In statistical mechanics, a master equation method is developed for the
calculation of the MFPT of random-walk problems \cite{1}, and there are many
researches about the master equation and its applications in the literature
\cite{KR,Gorban,Srokowski}. In this paper we will apply such a method to solve
the MFPT in quantum mechanics. For completeness, in this section, we first
give a detailed review for the method of the calculation of the MFPT through
solving a master equation, following Ref. \cite{1}. It will be seen that this
method can be used to solve the problem of MFPT in quantum mechanics without
any changes.

Consider a quantum system with $M$ states, $\left\vert 1\right\rangle $,
$\cdots$, $\left\vert M\right\rangle $. The system will transit among these
states under the interaction with the environment. In the following, we
calculate the MFPT of a transition starting from a given initial state
$\left\vert n_{0}\right\rangle $ at time $t=0$, going through some
intermediate states, to a given final state $\left\vert n_{p}\right\rangle $.

Let $P\left(  n,t\right)  $ be the probability that the system is at the state
$\left\vert n\right\rangle $ at time $t$, and $\Gamma_{mn}$ be the transition
probability rate of the transition from $\left\vert n\right\rangle $ to
$\left\vert m\right\rangle $. Then, in time interval $\Delta t$, the
transition probability of the transition from $\left\vert n\right\rangle $ to
$\left\vert m\right\rangle $ is $\Gamma_{mn}\Delta t$. The evolution of such a
system obeys the master equation,%
\begin{equation}
\frac{\partial}{\partial t}P\left(  n,t\right)  =%
{\displaystyle\sum\limits_{m=1}^{M}}
\left[  \Gamma_{nm}P\left(  m,t\right)  -\Gamma_{mn}P\left(  n,t\right)
\right]  . \label{masterEq}%
\end{equation}
The first part of the right-hand side of Eq. (\ref{masterEq}) represents the
transition probability from all other states to $\left\vert n\right\rangle $,
and the second part represents the transition probability from $\left\vert
n\right\rangle $ to other states.

The master equation can be written in a compact form by introducing a
transition matrix $W$ with matrix elements $W_{nm}=\Gamma_{nm}-\delta_{nm}%
{\displaystyle\sum\limits_{k=1}^{M}}
\Gamma_{kn}$ and a column vector $\left\vert P\left(  t\right)  \right\rangle
=\left(  P\left(  1,t\right)  ,P\left(  2,t\right)  ,\cdots,P\left(
M,t\right)  \right)  ^{T}$ denoting the state of the system at time $t$. Here,
the vector $\left\vert P\left(  t\right)  \right\rangle $ and the matrix $W$
satisfy $%
{\displaystyle\sum\limits_{n=1}^{M}}
P\left(  n,t\right)  =1$, $W _{\substack{nm \\\left(  n\neq m\right)  }}\geq
0$, and $%
{\displaystyle\sum\limits_{n=1}^{M}}
W_{nm}=0$ ($m=1,2,\cdots,M$). Then the master equation (\ref{masterEq})
becomes%
\begin{equation}
\frac{d}{dt}\left\vert P\left(  t\right)  \right\rangle =W\left\vert P\left(
t\right)  \right\rangle . \label{MEq}%
\end{equation}

For a system at the initial state $\left\vert n_{0}\right\rangle $ when $t=0$,
we now calculate the average time for the system transiting to a given final
state $\left\vert n_{p}\right\rangle $ for the first time, i.e., the MFPT.

Let $Q_{n}\left(  t\right)  $ denote the conditional probability for the
system being at state $\left\vert n\right\rangle $ at time $t$ given the
condition that this system is at state $\left\vert n_{0}\right\rangle $ when
$t=0$. $Q_{n}\left(  t\right)  $ satisfies the same master equation as
$P\left(  n,t\right)  $, Eq. (\ref{masterEq}), with the initial condition
$Q_{n}\left(  0\right)  =\delta_{nn_{0}}$. Take the absorbing boundary
condition $Q_{n_{p}}\left(  t\right)  =0$, then%
\begin{equation}
\frac{\partial}{\partial t}Q_{n}\left(  t\right)  =%
{\displaystyle\sum\limits_{m=1}^{M}}
\left[  \Gamma_{nm}Q_{m}\left(  t\right)  -\Gamma_{mn}Q_{n}\left(  t\right)
\right]  ,\text{ }n\neq n_{p}.
\end{equation}

Also, we can write this equation in the matrix form%
\begin{equation}
\frac{d}{dt}Q\left(  t\right)  =MQ\left(  t\right)  , \label{EqQ}%
\end{equation}
where the matrix element $Q_{n}\left(  t\right)  =\left\langle n\right\vert
Q\left(  t\right)  \left\vert n_{0}\right\rangle $ ($n\neq n_{p}$) and
$M_{nm}=\Gamma_{nm}-\delta_{nm}%
{\displaystyle\sum\limits_{k=1}^{M}}
\Gamma_{kn}$ ($n\neq n_{p}$ and $m\neq n_{p}$). Here, $M$ and $Q\left(
t\right)  $ are obtained by removing the $n_{p}$-th component from the matrix
$W$ and the vector $\left\vert P\left(  t\right)  \right\rangle $.

Using the method of the matrix spectral decomposition, we can solve $Q\left(
t\right)  $:%
\begin{equation}%
\begin{array}
[c]{l}%
Q\left(  t\right)  =Q\left(  0\right)
{\displaystyle\sum\limits_{i=1}^{M-1}}
e^{\lambda_{i}t}\left\vert \psi_{i}\right\rangle \left\langle \chi
_{i}\right\vert ,\\
\left\langle n\right\vert Q\left(  0\right)  \left\vert n_{0}\right\rangle
=\delta_{nn_{0}},
\end{array}
\end{equation}
where $\lambda_{i}$ is the $i$-th eigenvalue of the matrix $M$, $\left\langle
\chi_{i}\right\vert $ and $\left\vert \psi_{i}\right\rangle $ are the
normalized orthogonal left eigenvector and right eigenvector belonging to
$\lambda_{i}$. These eigenvectors satisfy%
\begin{align}
\left\langle \chi_{i}\right.  \left\vert \psi_{j}\right\rangle  &
=\delta_{ij},\nonumber\\%
{\displaystyle\sum\limits_{i=1}^{M-1}}
\left\vert \psi_{i}\right\rangle \left\langle \chi_{i}\right\vert  &  =I.
\end{align}
Then, $Q_{n}\left(  t\right)  $ reads%
\begin{equation}
Q_{n}\left(  t\right)  =\left\langle n\right\vert Q\left(  t\right)
\left\vert n_{0}\right\rangle =%
{\displaystyle\sum\limits_{i=1}^{M-1}}
e^{\lambda_{i}t}\left\langle n\right.  \left\vert \psi_{j}\right\rangle
\left\langle \chi_{i}\right.  \left\vert n_{0}\right\rangle ,\text{ }n\neq
n_{p}.
\end{equation}

Let $P\left(  t\right)  $ denote the probability that the system has not
transited to the state $\left\vert n_{p}\right\rangle $ at the time $t$; in
other words, $P\left(  t\right)  $ is the sum of the probability that the
system is at other states, i.e.,%
\begin{equation}
P\left(  t\right)  =%
{\displaystyle\sum\limits_{n=1,n\neq n_{p}}^{M}}
Q_{n}\left(  t\right)  .
\end{equation}

Introducing $f_{n_{p}}\left(  t\right)  dt$ to represent the probability that
the system transits to the state $\left\vert n_{p}\right\rangle $ during the
time interval $t\rightarrow t+dt$, we have%
\begin{equation}
f_{n_{p}}\left(  t\right)  =-\frac{d}{dt}P\left(  t\right)  .
\end{equation}
Therefore, the MFPT from $\left\vert n_{0}\right\rangle $ to $\left\vert
n_{p}\right\rangle $ is%
\begin{equation}
\left\langle t\right\rangle =\int_{0}^{\infty}tf_{n_{p}}\left(  t\right)
dt=\int_{0}^{\infty}P\left(  t\right)  dt.
\end{equation}

\section{Mean first-passage time of multiple-state systems}

In this section, we will calculate the mean first-passage time of a transition
from the highest state to the lowest state through spontaneous transitions for
three-, four-, and five-state systems, respectively.

\textit{Two-state systems}: Obviously, the MFPT of a transition from the
higher state to the lower state of a two-state system is just the lifetime of
the excited state, i.e.,%
\begin{equation}
\left\langle t\right\rangle =\frac{1}{\Gamma_{01}},
\end{equation}
where $\Gamma_{01}$ is the spontaneous transition probability rate of the
transition from the excited state $\left\vert 1\right\rangle $ to the ground
state $\left\vert 0\right\rangle $.

\textit{Three-state systems: }Consider a three-state system with the ground
state $\left\vert 0\right\rangle $, the first excited state $\left\vert
1\right\rangle $, and the second excited state $\left\vert 2\right\rangle $ .
Let $\Gamma_{02}$, $\Gamma_{01}$, and $\Gamma_{12}$ be the spontaneous
transition probability rates among these three states.

Suppose that at time $t=0$ the system is at state $\left\vert 2\right\rangle
$. Now, we calculate the MFPT for the transition $\left\vert 2\right\rangle
\rightarrow\left\vert 0\right\rangle $.

The state vector and the transition matrix are then%
\begin{equation}
\left\vert P\left(  t\right)  \right\rangle =%
\begin{pmatrix}
P\left(  0,t\right) \\
P\left(  1,t\right) \\
P\left(  2,t\right)
\end{pmatrix}
\label{19}%
\end{equation}
and%
\begin{equation}
W=%
\begin{pmatrix}
0 & \Gamma_{01} & \Gamma_{02}\\
0 & -\Gamma_{01} & \Gamma_{12}\\
0 & 0 & -\Gamma_{02}-\Gamma_{12}%
\end{pmatrix}
, \label{19.5}%
\end{equation}
respectively. Removing the zeroth component, we obtain the conditional
probabilities,%
\begin{align}
Q_{1}\left(  t\right)   &  =\left\langle 1\right\vert Q\left(  t\right)
\left\vert 2\right\rangle ,\\
Q_{2}\left(  t\right)   &  =\left\langle 2\right\vert Q\left(  t\right)
\left\vert 2\right\rangle ,\nonumber
\end{align}
and the matrix%
\begin{equation}
M=%
\begin{pmatrix}
-\Gamma_{01} & \Gamma_{12}\\
0 & -\Gamma_{02}-\Gamma_{12}%
\end{pmatrix}
.
\end{equation}
$Q\left(  t\right)  $ is determined by Eq. (\ref{EqQ}), i.e.,$\frac{d}%
{dt}Q\left(  t\right)  =MQ\left(  t\right)  $, with the initial condition%
\begin{equation}
\left\langle n\right\vert Q\left(  0\right)  \left\vert 2\right\rangle
=\delta_{n2}.
\end{equation}
Solving this equation directly, we achieve%
\begin{align}
Q_{1}\left(  t\right)   &  =\frac{\Gamma_{12}}{\Gamma_{02}+\Gamma_{12}%
-\Gamma_{01}}\left[  1-e^{-\left(  \Gamma_{02}+\Gamma_{12}-\Gamma_{01}\right)
t}\right]  e^{-\Gamma_{01}t},\nonumber\\
Q_{2}\left(  t\right)   &  =e^{-\left(  \Gamma_{02}+\Gamma_{12}\right)  t}.
\end{align}
Thus the MFPT of such a transition process is%
\begin{align}
\left\langle t\right\rangle  &  =\int_{0}^{\infty}\left[  Q_{1}\left(
t\right)  +Q_{2}\left(  t\right)  \right]  dt\nonumber\\
&  =\frac{\Gamma_{01}+\Gamma_{12}}{\Gamma_{01}\left(  \Gamma_{02}+\Gamma
_{12}\right)  }. \label{3state}%
\end{align}

\textit{Four-state systems:} By the same approach, we can calculate the MFPT
for a four-state system with the ground state $\left\vert 0\right\rangle $ and
the excited states $\left\vert 1\right\rangle $, $\left\vert 2\right\rangle $,
and $\left\vert 3\right\rangle $ with the spontaneous transition rates
$\Gamma_{03}$, $\Gamma_{02}$, $\Gamma_{01}$, $\Gamma_{13}$, $\Gamma_{12}$, and
$\Gamma_{23}$:%
\begin{equation}
\left\langle t\right\rangle _{\left\vert 3\right\rangle \rightarrow\left\vert
0\right\rangle }=\frac{\left(  \Gamma_{12}+\Gamma_{02}\right)  \left(
\Gamma_{01}+\Gamma_{13}\right)  +\Gamma_{23}\left(  \Gamma_{01}+\Gamma
_{12}\right)  }{\Gamma_{01}\left(  \Gamma_{02}+\Gamma_{12}\right)  \left(
\Gamma_{03}+\Gamma_{13}+\Gamma_{23}\right)  }.
\end{equation}

\textit{Five-state systems:} Similarly, for a five-state system, the MFPT is%
\begin{align}
&  \left\langle t\right\rangle _{_{\left\vert 4\right\rangle \rightarrow
\left\vert 0\right\rangle }}\nonumber\\
&  =\frac{\left(  \Gamma_{01}+\Gamma_{12}\right)  \left[  \Gamma_{23}%
\Gamma_{34}+\Gamma_{24}\left(  \Gamma_{03}+\Gamma_{13}+\Gamma_{23}\right)
\right]  }{\Gamma_{01}\left(  \Gamma_{02}+\Gamma_{12}\right)  \left(
\Gamma_{03}+\Gamma_{13}+\Gamma_{23}\right)  \left(  \Gamma_{04}+\Gamma
_{14}+\Gamma_{24}+\Gamma_{34}\right)  }\nonumber\\
&  +\frac{\Gamma_{34}\left(  \Gamma_{01}+\Gamma_{13}\right)  \left(
\Gamma_{12}+\Gamma_{02}\right)  +\left(  \Gamma_{01}+\Gamma_{14}\right)
\left(  \Gamma_{12}+\Gamma_{02}\right)  \left(  \Gamma_{03}+\Gamma_{13}%
+\Gamma_{23}\right)  }{\Gamma_{01}\left(  \Gamma_{02}+\Gamma_{12}\right)
\left(  \Gamma_{03}+\Gamma_{13}+\Gamma_{23}\right)  \left(  \Gamma_{04}%
+\Gamma_{14}+\Gamma_{24}+\Gamma_{34}\right)  }.
\end{align}

Once $\Gamma_{ij}=\Gamma=\mathrm{const}$ ($i<j$), we can obtain the MFPT of an
$n$-state system: $\left\langle t\right\rangle _{_{\left\vert n-1\right\rangle
\rightarrow\left\vert 0\right\rangle }}=1/\Gamma$. This is just like a
two-state system with the spontaneous transition rate $\Gamma$.

\section{Transitions with background interferences}

In this section, we consider a transition with background interferences.

A transition occurring in an environment background is an
important problem in practice. For example, in realistic quantum
information and quantum computation processes, the background
interference cannot be ignored \cite{BGA,Per,PDMF,CDPR,DLAFL}. In
the following, we will discuss the influence of environment
backgrounds on a transition process by examples.

First, consider a model that a transition process of a two-state system occurs
in an environment background. The two-state system consists of a ground state
$\left\vert g\right\rangle $ and an excited state $\left\vert e\right\rangle
$. The transition probability rate of the spontaneous transition from
$\left\vert e\right\rangle $ to $\left\vert g\right\rangle $ is $\Gamma$.
Suppose that the environment consists of $n$ states $\left\vert 1\right\rangle
$, $\left\vert 2\right\rangle $, ..., $\left\vert n\right\rangle $ satisfying
$E_{e}>E_{n}>E_{n-1}>\cdots>E_{1}>E_{g}$. The transition probability rate
between the background states $\left\vert i\right\rangle $ and $\left\vert
j\right\rangle $ is denoted as $\Gamma_{ij} $. For convenience, we denote the
transition probability rate of the spontaneous transition from $\left\vert
e\right\rangle $ to $\left\vert i\right\rangle $ by $\Gamma_{i,n+1}$ and from
$\left\vert i\right\rangle $ to $\left\vert g\right\rangle $ by $\Gamma_{0i}$.

If there is no influence of background interferences, the MFPT from
$\left\vert e\right\rangle $ to $\left\vert g\right\rangle $ is just the
lifetime of the excited state $\left\vert e\right\rangle $, i.e.,
$\tau=1/\Gamma$. If there exist background interferences, however, the MFPT
from $\left\vert e\right\rangle $ to $\left\vert g\right\rangle $ will be
influenced. In the following, we will consider two simplified models of
background interferences.

\textit{Case (1):} $\left\vert e\right\rangle \rightarrow\left\vert
n\right\rangle \rightarrow\left\vert n-1\right\rangle \rightarrow
\cdots\rightarrow\left\vert 1\right\rangle \rightarrow\left\vert
g\right\rangle $.

For a transition process $\left\vert e\right\rangle \rightarrow\left\vert
n\right\rangle \rightarrow\left\vert n-1\right\rangle \rightarrow
\cdots\rightarrow\left\vert 1\right\rangle \rightarrow\left\vert
g\right\rangle $, we can directly obtain the MFPT:%
\begin{align}
\left\langle t\right\rangle _{\left\vert e\right\rangle \rightarrow\left\vert
g\right\rangle }  &  =\frac{\Gamma_{n,n+1}}{\Gamma_{n,n+1}+\Gamma}\left(
\frac{1}{\Gamma_{01}}+\frac{1}{\Gamma_{12}}+\cdots+\frac{1}{\Gamma_{n,n+1}%
}\right) \nonumber\\
&  =\frac{\Gamma_{n,n+1}}{\Gamma_{n,n+1}+\Gamma}%
{\displaystyle\sum\limits_{k=1}^{n+1}}
\frac{1}{\Gamma_{k-1,k}}.
\end{align}

Consider a simple case: $\Gamma_{i,i+1}=\gamma=\mathrm{const}$. In such a
case, we have%
\begin{equation}
\left\langle t\right\rangle _{\left\vert e\right\rangle \rightarrow\left\vert
g\right\rangle }=\frac{n+1}{\gamma+\Gamma}. \label{tetog}%
\end{equation}
Obviously, the MFPT is proportional to the total number of the background states.

When $\gamma\gg\Gamma$, i.e., the transition probability rate of the
transition involving background states is much greater than the transition
probability rate between the two system states, the MFPT is%
\begin{equation}
\left\langle t\right\rangle _{\left\vert e\right\rangle \rightarrow\left\vert
g\right\rangle }\sim\frac{n+1}{\gamma}.
\end{equation}
In this case, when the influence of the background dominates, the MFPT is
almost determined by the transition probability among background states and
the total number of the background states.

When $\gamma\ll\Gamma$, the transition probability, Eq. (\ref{tetog}),
becomes
\begin{equation}
\left\langle t\right\rangle _{\left\vert e\right\rangle \rightarrow\left\vert
g\right\rangle }=\frac{n+1}{\Gamma}.
\end{equation}
In this case, the MFPT is determined by the transition probability between two
system states and the total number of the background states.

\textit{Case (2):} $\left\vert e\right\rangle \rightarrow\left\vert
\text{\textrm{background} \textrm{states}}\right\rangle \rightarrow\left\vert
g\right\rangle $ with $\left\vert \text{\textrm{background} \textrm{states}%
}\right\rangle $ denoting $n$ intermediate states.

In this case, the background consists of $n$ background states. Suppose that
the transition probability rates from the excited state $\left\vert
e\right\rangle $ to every background states are the same, denoted by
$\Gamma_{1}$, and the transition probability rates from every background
states to the ground state $\left\vert g\right\rangle $ are the same, denoted
by $\Gamma_{2}$. Moreover, we also suppose that the transition probability
among the background states is zero. We then can obtain the MFPT from
$\left\vert e\right\rangle $ to $\left\vert g\right\rangle $:%
\begin{equation}
\left\langle t\right\rangle _{\left\vert e\right\rangle \rightarrow\left\vert
g\right\rangle }=\frac{n\Gamma_{1}+\Gamma_{2}}{\Gamma_{2}\left(  n\Gamma
_{1}+\Gamma\right)  }. \label{eng}%
\end{equation}
Comparing with Eq. (\ref{3state}), we can see that such a transition is just
the transition among three states with $\Gamma_{12}$ replaced by $n\Gamma_{1}
$.

When $\Gamma_{i}$ $\gg\Gamma$, the MFPT, Eq. (\ref{eng}), becomes%
\begin{equation}
\left\langle t\right\rangle _{\left\vert e\right\rangle \rightarrow\left\vert
g\right\rangle }\sim\frac{1}{n\Gamma_{1}}+\frac{1}{\Gamma_{2}}.
\end{equation}
The MFPT is almost independent of the transition probability between the
system states. Moreover, if $\Gamma_{i}\gg\Gamma$ and $n\gg1$, we arrive at%
\begin{equation}
\left\langle t\right\rangle _{\left\vert e\right\rangle \rightarrow\left\vert
g\right\rangle }\sim\frac{1}{\Gamma_{2}},
\end{equation}
i.e., the MFPT is only determined by the transition probability from the
background states to the ground state $\left\vert g\right\rangle $.

In another case, when $n$ is not large enough so that $n\Gamma_{1}\ll\Gamma$,
$\left\langle t\right\rangle \sim\left(  1+n\Gamma_{1}/\Gamma_{2}\right)
/\Gamma.$ That is different from the intuitive result $1/\Gamma$, but relies
on $n\Gamma_{1}/\Gamma_{2}$.

\section{Roundabout transitions: hydrogen atoms}

In this section, we consider a roundabout transition process in a hydrogen atom.

Selection rules forbid certain transitions in a hydrogen atom. Nevertheless,
what the selection rule forbids is the direct transition, a transition between
two states. If a transition between two states $\left\vert i\right\rangle $
and $\left\vert f\right\rangle $ is forbidden, however, a roundabout
transition from $\left\vert i\right\rangle $ to $\left\vert f\right\rangle $
can also occur.

Concretely, even the transition $\left\vert i\right\rangle \rightarrow
\left\vert f\right\rangle $ is forbidden, a roundabout transition $\left\vert
i\right\rangle \rightarrow\left\vert \text{\textrm{intermediate}
\textrm{states}}\right\rangle \rightarrow\left\vert f\right\rangle $ can in
principle occur through some intermediate states, $\left\vert n\right\rangle
$, $\left\vert n-1\right\rangle $, $\cdots$, $\left\vert 1\right\rangle $,
between $\left\vert i\right\rangle $ and $\left\vert f\right\rangle $.

In this paper, taking hydrogen atoms as an example, we calculate the MFPT for
such a roundabout transition.

For hydrogen atoms, the selection rule of electric dipole transition between
states $\left\vert nlm\right\rangle $ read%
\begin{equation}
\Delta l=\pm1,\text{ }\Delta m=0,\pm1,
\end{equation}
where $n$, $l$, and $m$ are principal quantum number, angular quantum number,
and magnetic quantum number, respectively.

In the following, as an example, we calculate the MFPT of a roundabout
transition from $\left\vert 300\right\rangle $ to $\left\vert 100\right\rangle
$; the direct transition from $\left\vert 300\right\rangle $ to $\left\vert
100\right\rangle $ is forbidden by the selection rule of electric dipole
transition. Nevertheless, there still some roundabout transitions from
$\left\vert 300\right\rangle $ to $\left\vert 100\right\rangle $ can occur:%
\begin{equation}
\left\vert 300\right\rangle \rightarrow\left\vert 21m\right\rangle
\rightarrow\left\vert 100\right\rangle ,\text{ }m=1,0,-1. \label{roundabout}%
\end{equation}

In this case, the corresponding vector $\left\vert P\left(  t\right)
\right\rangle $ and transition matrix $W$ are%
\begin{equation}
\left\vert P\left(  t\right)  \right\rangle =%
\begin{pmatrix}
P\left(  \left\vert 100\right\rangle ,t\right) \\
P\left(  \left\vert 21,-1\right\rangle ,t\right) \\
P\left(  \left\vert 210\right\rangle ,t\right) \\
P\left(  \left\vert 211\right\rangle ,t\right) \\
P\left(  \left\vert 300\right\rangle ,t\right)
\end{pmatrix}
,
\end{equation}
and%

\begin{equation}
W=%
\begin{pmatrix}
0 & \Gamma_{\left\vert 21,-1\right\rangle \rightarrow\left\vert
100\right\rangle } & \Gamma_{\left\vert 210\right\rangle \rightarrow\left\vert
100\right\rangle } & \Gamma_{\left\vert 211\right\rangle \rightarrow\left\vert
100\right\rangle } & 0\\
0 & -\Gamma_{\left\vert 21,-1\right\rangle \rightarrow\left\vert
100\right\rangle } & 0 & 0 & \Gamma_{\left\vert 300\right\rangle
\rightarrow\left\vert 21,-1\right\rangle }\\
0 & 0 & -\Gamma_{\left\vert 210\right\rangle \rightarrow\left\vert
100\right\rangle } & 0 & \Gamma_{\left\vert 300\right\rangle \rightarrow
\left\vert 210\right\rangle }\\
0 & 0 & 0 & -\Gamma_{\left\vert 211\right\rangle \rightarrow\left\vert
100\right\rangle } & \Gamma_{\left\vert 300\right\rangle \rightarrow\left\vert
211\right\rangle }\\
0 & 0 & 0 & 0 & -\Gamma_{\left\vert 300\right\rangle \rightarrow\left\vert
21,-1\right\rangle }-\Gamma_{\left\vert 300\right\rangle \rightarrow\left\vert
210\right\rangle }-\Gamma_{\left\vert 300\right\rangle \rightarrow\left\vert
211\right\rangle }%
\end{pmatrix}
.
\end{equation}

The master equation given by Eq. (\ref{MEq}) is $\frac{d}{dt}\left\vert
P\left(  t\right)  \right\rangle =W\left\vert P\left(  t\right)  \right\rangle
$.

In this problem, the initial state is $\left\vert 300\right\rangle $ and the
final state is $\left\vert 100\right\rangle $, so $Q\left(  t\right)  $ and
$M$ read%
\begin{align}
Q_{\left\vert 21,-1\right\rangle }\left(  t\right)   &  =\left\langle
21,-1\right\vert Q\left(  t\right)  \left\vert 300\right\rangle ,\nonumber\\
Q_{\left\vert 210\right\rangle }\left(  t\right)   &  =\left\langle
210\right\vert Q\left(  t\right)  \left\vert 300\right\rangle ,\nonumber\\
Q_{\left\vert 211\right\rangle }\left(  t\right)   &  =\left\langle
211\right\vert Q\left(  t\right)  \left\vert 300\right\rangle ,\nonumber\\
Q_{\left\vert 300\right\rangle }\left(  t\right)   &  =\left\langle
300\right\vert Q\left(  t\right)  \left\vert 300\right\rangle
\end{align}
and%

\begin{equation}
M=%
\begin{pmatrix}
-\Gamma_{\left\vert 21,-1\right\rangle \rightarrow\left\vert 100\right\rangle
} & 0 & 0 & \Gamma_{\left\vert 300\right\rangle \rightarrow\left\vert
21,-1\right\rangle }\\
0 & -\Gamma_{\left\vert 210\right\rangle \rightarrow\left\vert
100\right\rangle } & 0 & \Gamma_{\left\vert 300\right\rangle \rightarrow
\left\vert 210\right\rangle }\\
0 & 0 & -\Gamma_{\left\vert 211\right\rangle \rightarrow\left\vert
100\right\rangle } & \Gamma_{\left\vert 300\right\rangle \rightarrow\left\vert
211\right\rangle }\\
0 & 0 & 0 & -\Gamma_{\left\vert 300\right\rangle \rightarrow\left\vert
21,-1\right\rangle }-\Gamma_{\left\vert 300\right\rangle \rightarrow\left\vert
210\right\rangle }-\Gamma_{\left\vert 300\right\rangle \rightarrow\left\vert
211\right\rangle }%
\end{pmatrix}
.
\end{equation}
Here, $Q\left(  t\right)  $ is determined by $\frac{d}{dt}Q\left(  t\right)
=MQ\left(  t\right)  $ with the initial condition%
\begin{equation}
\left\langle 300\right\vert Q\left(  0\right)  \left\vert 300\right\rangle
=1\text{ and }\left\langle 21m\right\vert Q\left(  0\right)  \left\vert
300\right\rangle =0.
\end{equation}

Then, by solving $Q\left(  t\right)  $, we achieve the MFPT of the roundabout
transition (\ref{roundabout}):%
\begin{equation}
\left\langle t\right\rangle =\int_{0}^{\infty}\left[  Q_{\left\vert
21,-1\right\rangle }\left(  t\right)  +Q_{\left\vert 210\right\rangle }\left(
t\right)  +Q_{\left\vert 211\right\rangle }\left(  t\right)  +Q_{\left\vert
300\right\rangle }\left(  t\right)  \right]  dt.
\end{equation}
Here, $\Gamma_{\left\vert 300\right\rangle \rightarrow\left\vert
21m\right\rangle }=2.1046\times10^{6}s^{-1}$ and $\Gamma_{\left\vert
21m\right\rangle \rightarrow\left\vert 100\right\rangle }=6.2649\times
10^{8}s^{-1}$ ($m=0,\pm1$) \cite{WF}, we then have
\begin{equation}
\left\langle t\right\rangle _{\left\vert 300\right\rangle \rightarrow
\left\vert 21m\right\rangle \rightarrow\left\vert 100\right\rangle
}=1.5998\times10^{-7}s. \label{Hatom}%
\end{equation}

It should be noted here that the MFPT, $\left\langle t\right\rangle $, is
different from the lifetime of a state. The lifetime does not associate a
certain final state. The MFPT, however, associates a certain initial state and
a certain final state. In this case, the lifetime of $\left\vert
300\right\rangle $ is $1.5838\times10^{-7}s$, which is very close to the value
of the MFPT, $\left\langle t\right\rangle $; while the lifetimes of
$\left\vert 21,-1\right\rangle $, $\left\vert 210\right\rangle $ and
$\left\vert 211\right\rangle $ are all $1.5962\times10^{-9}s$, which is so
small compared to the lifetime of $\left\vert 300\right\rangle $ and
$\left\langle t\right\rangle $. Thus, the transition from $\left\vert
300\right\rangle $ to $\left\vert 21,-1\right\rangle $, $\left\vert
210\right\rangle $ and $\left\vert 211\right\rangle $ is the dominant
contribution to the transition $\left\vert 300\right\rangle \rightarrow
\left\vert 21m\right\rangle \rightarrow\left\vert 100\right\rangle $.

Note that in this case, by chance, $\Gamma_{\left\vert 300\right\rangle
\rightarrow\left\vert 21-1\right\rangle }=\Gamma_{\left\vert 300\right\rangle
\rightarrow\left\vert 210\right\rangle }=\Gamma_{\left\vert 300\right\rangle
\rightarrow\left\vert 211\right\rangle }$, so we can also use the result
(\ref{eng}) to achieve Eq. (\ref{Hatom}) by only regarding the three states
$\left\vert 21,-1\right\rangle $, $\left\vert 210\right\rangle $, and
$\left\vert 211\right\rangle $ as background states and $\left\vert
300\right\rangle $ and $\left\vert 100\right\rangle $ as two system states.

\section{Applications to lasers}

The above result of the MFPT of a three-state system can be directly applied
to the problem of laser.

For a three-state laser scheme, let $\left\vert 0\right\rangle $, $\left\vert
1\right\rangle $, and $\left\vert 2\right\rangle $ represent the ground state,
the upper laser state, and the pumping state, respectively. The laser
procedure is realized as follows. Some pumping processes take atoms from
$\left\vert 0\right\rangle $ to $\left\vert 2\right\rangle $. The atoms at
$\left\vert 2\right\rangle $ drop very rapidly to the upper laser state
$\left\vert 1\right\rangle $, and the transition from $\left\vert
1\right\rangle $ to $\left\vert 0\right\rangle $ produces a photon we needed.
As long as the pumping process is effective enough, the lifetime of state
$\left\vert 2\right\rangle $ is short enough, and the lifetime of state
$\left\vert 1\right\rangle $ is long enough, the number of atoms in
$\left\vert 1\right\rangle $ will exceed the number of atoms in $\left\vert
0\right\rangle $, i.e., the population inversion will be achieved. The
transition between $\left\vert 1\right\rangle $ to $\left\vert 0\right\rangle
$ will yield a laser. According to the above discussion, if we want to obtain
a stable output of laser, a necessary condition is that the probability of an
atom pumping from $\left\vert 0\right\rangle $ to $\left\vert 2\right\rangle $
must exceed the probability of an atom transiting from $\left\vert
2\right\rangle $ to $\left\vert 0\right\rangle $.

As the above discussion, the MFPT of the transition $\left\vert 2\right\rangle
$ to $\left\vert 0\right\rangle $ can describe the probability of the
transition from $\left\vert 2\right\rangle $ to $\left\vert 0\right\rangle $
exactly. If we introduce the pumping rate $P$ to describe the probability of
an atom pumping from $\left\vert 0\right\rangle $ to $\left\vert
2\right\rangle $, then the above necessary condition can be expressed as%
\begin{equation}
P\geq\frac{1}{\left\langle t\right\rangle },
\end{equation}
or
\begin{equation}
P\geq\Gamma_{01}\frac{\Gamma_{02}+\Gamma_{12}}{\Gamma_{01}+\Gamma_{12}}.
\label{PMFPT}%
\end{equation}

As a comparison, in the usual treatment of laser, the transition from
$\left\vert 2\right\rangle $ to $\left\vert 1\right\rangle $ is assumed as
instantaneous and the transition from $\left\vert 2\right\rangle $ to
$\left\vert 0\right\rangle $ is neglected, which means%
\begin{equation}
\Gamma_{12}\rightarrow\infty\text{ and }\Gamma_{02}\rightarrow0. \label{app}%
\end{equation}
Under these assumptions, the usual result of the pumping rate satisfies
\cite{ME}%
\begin{equation}
P\geq\Gamma_{01}. \label{P0}%
\end{equation}

From Eq. (\ref{PMFPT}), we can see that our result based on the MFPT will
reduce to the usual result (\ref{P0}) under the condition Eq. (\ref{app}); Eq.
(\ref{PMFPT}) is a more accurate result. In fact, the power of a laser is
usually expressed as \cite{ME,EPJ}
\begin{equation}
P_{\mathrm{laser}}=A\left(  P-\Gamma_{01}\right)  ,
\end{equation}
where $A$ is a parameter determined by the character of the specific laser
medium and laser facility. However, a more accurate expression of the power is
expressed by the MFPT:%
\begin{equation}
P_{\mathrm{laser}}=A\left(  P-\frac{1}{\left\langle t\right\rangle }\right)
=A\left(  P-\Gamma_{01}\frac{\Gamma_{02}+\Gamma_{12}}{\Gamma_{12}+\Gamma_{01}%
}\right)  .
\end{equation}

\section{Conclusions and outlook}

In this paper, we discuss the problem of the MFPT in quantum mechanics. In the
problem of the MFPT, we concentrate on a given transition process: $\left\vert
i\right\rangle \rightarrow\left\vert \text{\textrm{intermediate}
\textrm{states}}\right\rangle \rightarrow\left\vert f\right\rangle $.

We apply the method developed in statistical mechanics for calculating the
MFPT of the problem of random walks to calculate the MFPT of transition
processes in quantum mechanics. Such a method is based on the master equation.
Furthermore, we calculate the MFPT in multiple-state systems. Especially, we
consider a transition process occurring in an environment background by
examples. Taking a hydrogen atom as an example, we calculate a roundabout
transition process $\left\vert 300\right\rangle \mathbf{\rightarrow}\left\vert
100\right\rangle $. Finally, we discuss the application to laser theory.

When realizing a quantum information or a quantum computation
process in experiments, one has to face the influence of
environments. The environmental background interferences will
cause the problem of quantum decoherence
\cite{BGA,Per,PDMF,CDPR,DLAFL}. That is to say, in a realistic
quantum information or a quantum computation process, the
environmental background interferences cannot be ignored. Although
many researches are devoted to suppress the influence of
environment \cite{DLDVT,CFL,BPC,HF,CD}, decoherence is still one
of the most important obstacles in quantum information processes.
Based on the result of the present paper, we can view the
background interference as some intermediate states and consider
the quantum decoherence by analyzing the MFPT. We will discuss the
application to the problem of quantum decoherence elsewhere.

\vskip 0.5cm \noindent\textbf{Acknowledgements} This work is supported in part
by NSF of China under Grant No. 11075115.


\begin{thebibliography}{99}                                                                                               %


\bibitem {1}L.E. Reichl, A Modern Course in Statistical Physics, 2nd edn,
Wiley, New York, 1998.

\bibitem {2}C.W. Gardiner, Handbook of stochastic methods: for Physics,
Chemistry and the Natural Sciences, 3rd edn, Springer, Berlin, 2004.

\bibitem {CWKO}J.-L. Chen, C.F. Wu, L.C. Kwek, C.H. Oh, Phys. Rev. Lett. 93
(2004) 140407.

\bibitem {DWCO}D.-L. Deng, C.F. Wu, J.-L. Chen, C.H. Oh, Phys. Rev. Lett. 105
(2010) 060402.

\bibitem {BGA}J. Bergli, Y.M. Galperin, B.L. Altshuler, New J. Phys. 11 (2009) 025002.

\bibitem {Per}A. P\'{e}rez, Phys. Rev. A 81 (2010) 052326.

\bibitem {DLDVT}S. Damodarakurup, M. Lucamarini, G. Di Giuseppe, D. Vitali, P.
Tombesi, Phys. Rev. Lett. 103 (2009) 040502.

\bibitem {CFL}M. Castagnino, S. Fortin, O. Lombardi, J. Phys. A: Math. Theor.
43 (2010) 065304.

\bibitem {5}K. Lindenberg, K.E. Shuler, J. Freeman, T.J. Lie, J. Stat. Phys.
12 (1975) 217.

\bibitem {6}J.M. Sancho, Phys. Rev. A 31 (1985) 3523.

\bibitem {7}P. Hanggi, P. Talkner, Z. Phys. B 45 (1981) 79.

\bibitem {9}I. Klik, L. Gunther, J. Stat. Phys. 60 (1990) 473.

\bibitem {10}H. Hofmann, A.G. Magner, Phys. Rev. C 68 (2003) 014606.

\bibitem {13}D.C. Mei, G.Z. Xie, L. Cao, D.J. Wu, Phys. Rev. E 59 (1999) 3880.

\bibitem {15}U. Behn, Phys. Rev. E 47 (1993) 3970.

\bibitem {17}L. Cao, D.J. Wu, Phys. Lett. A 283 (2001) 313.

\bibitem {18}Z. Zhang, Y. Lin, S. Zhou, B. Wu, J. Guan, New J. Phys. 11 (2009) 103043.

\bibitem {19}E. Agliari, Phys. Rev. E 77 (2008) 011128.

\bibitem {21}B. Dybiec, E. Gudowska-Nowak, Phys. Rev. E 73 (2006) 046104.

\bibitem {23}P. Le Doussal, Phys. Rev. Lett. 62 (1989) 3097.

\bibitem {24}M.F. Shlesinger, Nature 450 (2007) 40.

\bibitem {27}A. Ansari, J. Chem. Phys. 112 (2000) 2516.

\bibitem {28}N.-V. Buchete, J. E. Straub, J. Phys. Chem. B 105 (2001) 6684.

\bibitem {KR}A. Kossakowski, R. Rebolledo, Open Sys. Infor. Dyn. 16 (2009) 259.

\bibitem {Gorban}A.N. Gorban, Physica A 390 (2011) 1009.

\bibitem {Srokowski}T. Srokowski, Physica A 390 (2011) 3077.

\bibitem {PDMF}E. Paladino, A. D'Arrigo, A. Mastellone, G. Falci, New J. Phys.
13 (2011) 093037.

\bibitem {CDPR}M.A. Cirone, G. De Chiara, G.M. Palma, A. Recati, New J. Phys.
11 (2009) 103055.

\bibitem {DLAFL}V. D'Auria, N. Lee, T. Amri, C. Fabre, J. Laurat, Phys. Rev.
Lett. 107 (2011) 050504.

\bibitem {WF}W.L. Wiese, J.R. Fuhr, J. Phys. Chem. Ref. Data 38 (2009) 565.

\bibitem {ME}P.W. Milonni, J.H. Eberly, Laser physics, Wiley, Hoboken, New
Jersey, 2010.

\bibitem {EPJ}W.-S. Dai, M. Xie, Eur. Phys. J. D 19 (2002) 125.

\bibitem {BPC}P.G. Brooke, M.K. Patra, J.D. Cresser, Phys. Rev. A 77 (2008) 062313.

\bibitem {HF}J.-T. Hsiang, L.H. Ford, Int. J. Mod. Phys. A 24 (2009) 1705.

\bibitem {CD}R. Chaves, L. Davidovich, Phys. Rev. A 82 (2010) 052308.
\end{thebibliography}
\end{document}